\documentclass[11pt]{amsart}

\usepackage{amsfonts}
\usepackage{amsmath}
\usepackage{hyperref}
\usepackage{rotating}
\usepackage{color}
\numberwithin{equation}{section}

\newcommand{\e}{\mathrm{e}}
\newcommand{\dd}{\mathrm{d}}
\newcommand{\minu}{\! - \!}

\begin{document}

\title[Territorial Behaviour of Buzzards versus Random Matrix Spacing
Distributions] {\textbf{Territorial Behaviour of Buzzards versus\\[2mm]
Random Matrix Spacing Distributions }}
	
\author[G.~Akemann et al.]{Gernot Akemann, Michael Baake, Nayden
  Chakarov,  Oliver Kr\"{u}ger,\\  {\vspace*{-2mm}}  \\
  Adam Mielke, Meinolf Ottensmann, and Rebecca Werdehausen}

\address{Faculty of Biology (NC, OK, MO),
  Faculty of Mathematics (MB),
  \newline \indent and Faculty of Physics (GA, AM, RW),
  Bielefeld University, \newline \indent
  P.O. Box 100131, D-33501 Bielefeld, Germany}
\email{akemann@physik.uni-bielefeld.de, mbaake@math.uni-bielefeld.de,
  \newline \indent oliver.krueger@uni-bielefeld.de,
  amielke@math.uni-bielefeld.de}

\begin{abstract}
  A deeper understanding of the processes underlying the distribution
  of animals in space is crucial for both basic and applied
  ecology. The Common buzzard (Buteo buteo) is a highly aggressive,
  territorial bird of prey that interacts strongly with its intra- and
  interspecific competitors. We propose and use random matrix theory
  to quantify the strength and range of repulsion as a function of the
  buzzard population density, thus providing a novel approach to model
  density dependence. As an indicator of territorial behaviour, we
  perform a large-scale analysis of the distribution of buzzard nests
  in an area of $300$ square kilometres around the Teutoburger Wald,
  Germany, as gathered over a period of $20$ years. The nearest and
  next-to-nearest neighbour spacing distribution between nests is
  compared to the two-dimensional Poisson distribution, originating
  from uncorrelated random variables, to the complex eigenvalues of
  random matrices, which are strongly correlated, and to a
  two-dimensional Coulomb gas interpolating between these two. A
  one-parameter fit to a time-moving average reveals a significant
  increase of repulsion between neighbouring nests, as a function of
  the observed increase in absolute population density over the
  monitored period of time, thereby proving an unexpected yet simple
  model for density-dependent spacing of predator territories. A
  similar effect is obtained for next-to-nearest neighbours, albeit
  with weaker repulsion, indicating a short-range interaction. Our
  results show that random matrix theory might be useful in the
  context of population ecology.
\end{abstract}
 
\date{November 3, 2020}
\maketitle

%%%%%%%%%%%%%%%%%%%%%%%%%%%%%%
\section{Introduction} \label{Sec:Intro}

It is one of the major goals of population ecology, and indeed one of
the oldest goals of ecology as a whole, to understand the processes
that govern the distribution of animals in space and time~\cite{Lack}.
This is not only an important academic question, but has crucial
implications for conservation planning and management
\cite{KrC,OBrian}. Birds of prey, or raptors, are often apex
predators, are among the most-threatened groups of birds globally
\cite{KR}, and their abundance and diversity is a key indicator of the
state of the ecosystem as a whole \cite{Ser}.  Birds of prey have been
used as model systems in other such contexts, for example vultures,
including the interaction with humans \cite{Gangoso}, or even as an
indicator for the well-being of entire ecosystems \cite{KrC,OBrian}.
Their charismatic nature, conspicuousness and relatively high
vulnerability due to their trophic position increases their
conservation value \cite{Ser} and hence the recording of their
population dynamics has attracted a lot of attention for centuries
\cite{New}.  In this article, we focus on the behaviour of such a
territorial bird of prey, the common buzzard, and model a change in
its population density over $20$ years.  As we shall see, their
territoriality, measured by the repulsion among their nests, increases
significantly with population size.

In many animal and plant populations, reproductive success decreases
with increasing population density. This density dependence of
reproduction has been known since the dawn of modern animal
ecology~\cite{Lack}. Although density dependence has been examined in
different taxa~\cite{SBDHP}, it is rather easily studied in large,
territorial species. One particularly prominent group used for
disentangling hypotheses about density-dependent processes have been
birds of prey~\cite{KrC,KL}. Their predatory habit could amplify the
occurrence of strong density dependence and make them especially
suited for studies of the underlying
mechanisms~\cite{DRP}. Territoriality allows density effects to be
examined in detail, while this possibility could be impaired in
classically colonial species.  Birds of prey also offer the advantage
of being very site-faithful: once they have occupied a territory, they
rarely move and they are highly aggressive against intruders and hence
territorial aggression can be fatal. The study of mechanisms that
could explain the spatial clustering of bird of prey territories is
therefore of theoretical as well as applied value.

Here, the locations of buzzard nests collected over the $20$ years
$2000 \minu 2019$ in the Teutoburger Wald around Bielefeld, Germany,
are investigated.  We are in the comfortable position to have about
$100 - 200$ nests per year available in this approximately two
dimensional (2D) landscape, over a period of $20$ years; see
Section~\ref{Sec:Buzz} for details.  Given such a data set, it is
tempting to model the emergence and behaviour in space and time from
an ecological angle. However, this would require to make many
assumptions and to introduce even more parameters, which bears the
danger of overfitting. Indeed, a less biased approach would begin by
extracting structural properties from the data, such as distribution
patterns, distance preferences, or any kinds of correlations in space
and time. Although the inference point of view from point process
theory would be natural, compare \cite{Karr}, the data set does not
seem to be large enough for such an endeavour.  Consequently, the goal
of this initial approach is rather modest in the sense that we
primarily look at the spacing distribution between neighbouring
points, that is, the nest locations, in order to test whether a model
with one parameter and no a priori biological assumptions could still
explain the change in territorial spacing over time.

A popular approach to spacing distributions is based on random matrix
ensembles.  They allow to analytically quantify the repulsion among
data points, without making further assumptions. In particular, the
spacing distributions of the classical random matrix ensembles in one
(1D) and two dimensions (2D) are parameter free. For some details on
random matrices, we refer to Appendix~\ref{Sec:Coulomb}.  Their
history started independently in multivariate statistics
\cite{Wishart}, inspired from agriculture by Wishart, and in a
statistical theory of energy levels of complex quantum systems by
Dyson \cite{Dyson}, motivated by Wigner and ideas of Bohr about the
compound nucleus; see \cite{GMW} for a historic account and various
modern applications.  Further examples without quantum mechanical
background repeatedly show the signature of random matrix statistics,
such as the spacing between subsequent buses in Cuernavaca
(Mexico)~\cite{Cuernavaca}, and parked cars \cite{RR,AM} 
or birds on a power line \cite{Seba}. 
While most comparisons are made for data in 1D, comparing
with the statistics of real eigenvalues of symmetric or Hermitian
random matrices, few examples exist in 2D. Here, one is comparing with
complex eigenvalues of random matrices without symmetries, with
applications ranging from quantum chaotic systems with dissipation
\cite{GHS1988,AKMP} over quantum field theory with chemical potential
\cite{Tilo} to the spacing between chief towns of departments or
districts \cite{LC1} and Swedish pine trees \cite{LC2}. Data sets
similar to the latter two have been modelled by so-called
determinantal point processes \cite{Moller}, for which random matrices
provide a particular example.  A novel feature we propose here is to
track the time evolution of the repulsion strength, in this case as a
function of the population density. This allows us to draw
biologically more relevant conclusions, compared to the static
distribution of birds on a power line. \smallskip

In order to apply such ideas to the distribution of nests, we employ
recent progress on the universality of certain distributions in random
matrix theory. In Section \ref{Sec:beta}, we start with a comparison
to the distribution from a uniform Poisson process in 2D, which
describes the distribution of uncorrelated points, and to the
distribution of the complex Ginibre ensemble of random matrices
\cite{Ginibre}, which displays a rather strong repulsion.  The
repulsive nature is also easily detectable from the diffuse scattering
components in the diffraction image of random point sets in 2D
\cite{BK}.  We stress again that both distributions are parameter
free, after fixing the normalisation and first moment to unity.  As we
shall see, the nest locations are indeed not adequately described by a
Poisson process, in line with the known and frequently observed
territorial repulsion of the buzzards.

In this first step, it also becomes clear that the repulsion in the
Ginibre process is too strong, which is perhaps not too surprising
either, as the ecological system should show some repulsion on a
shorter scale (visual range), but not a long-range one. To deal with
this situation, we embark on a simple one-parameter interpolation
between Poisson and Ginibre statistics, which we derive from a known
2D Coulomb gas ensemble at variable temperature (being a
non-determinantal, general Gibbs point process \cite{For,Sylvia}).
Since the underlying model has no direct meaning in the biological
system, the parameter $\beta$ (proportional to the inverse
temperature) is just taken as an \textit{effective} phenomenological
quantity and then determined by a simple fitting procedure. It
directly measures the power of local repulsion of two points at
distance $s$, which is proportional to $s^{\beta+1}$.

It turns out that the employed one-parameter family of spacing
distributions works well for the data. Moreover, our effective
parameter proves sensitive to population density dependent properties
in time, which indicates its suitability for our initial step in the
data analysis. In Section \ref{Sec:beta}, we explain how we analyse
our data in moving time averages, with further details provided in
Appendix~\ref{C:Fit}.  Our conclusions and open questions are
summarised in Section~\ref{Sec:conc}.

%%%%%%%%%%%%%%%%%%%%%%%%%%%%%%%%%%%%%%

\section{Object of Study, Density Dependence, and 
Territoriality}\label{Sec:Buzz}

The Common buzzard (Buteo buteo L.) is a medium-sized bird of prey
($50 \minu 57$ cm body length, $525 \minu 1364$ g body weight) and
breeds across the Palaearctic~\cite{delHoyo1994}. Its main prey
consists of microtine rodents.  A Common buzzard population comprising
between 63 and 266 breeding pairs per year was monitored from $2000$
to $2019$ in an investigation area in Germany. The $300$ km$^2$ area
(8$^\circ$25'E and 52$^\circ$6'N) is located in Eastern Westphalia and
consists of two $125$ km$^2$ grid squares and $50$ km$^2$ edge areas.
The main habitat is the Teutoburger Wald, a low mountain region
reaching a height of $315$ m above sea level. Ridges are covered by
Norway Spruce Picea abies and Beech Fagus sylvatica, with Oak Quercus
robur and Q.~petrea forests at lower altitudes. The second main
habitat is a cultivated landscape to the north and south. In the
north, forests are composed mainly of Beech and Oak, whereas Scots
Pine Pinus sylvestris dominates in the south. The size of forest
patches varies from rows of trees to large patches of more than $10$
km$^2$ in size, and about $17\%$ of the study site is forested. Most
forests are less than 100 years old and the spruce forests are
atypical of this region. This study site has been intensively
monitored for Common buzzard and the resulting spatial data have been
used extensively before~\cite{KrC,CBK,MCHK}.  We use the very same
data of the nest locations for our analysis, completed by the most
recent years.

\begin{center}
\begin{figure}[ht]
  \includegraphics[width=0.45\linewidth,angle=0]{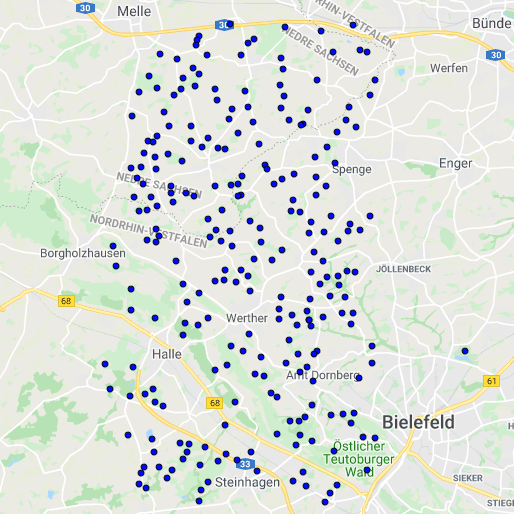}
  \hspace{10pt}
  \includegraphics[width=0.45\linewidth,angle=0]{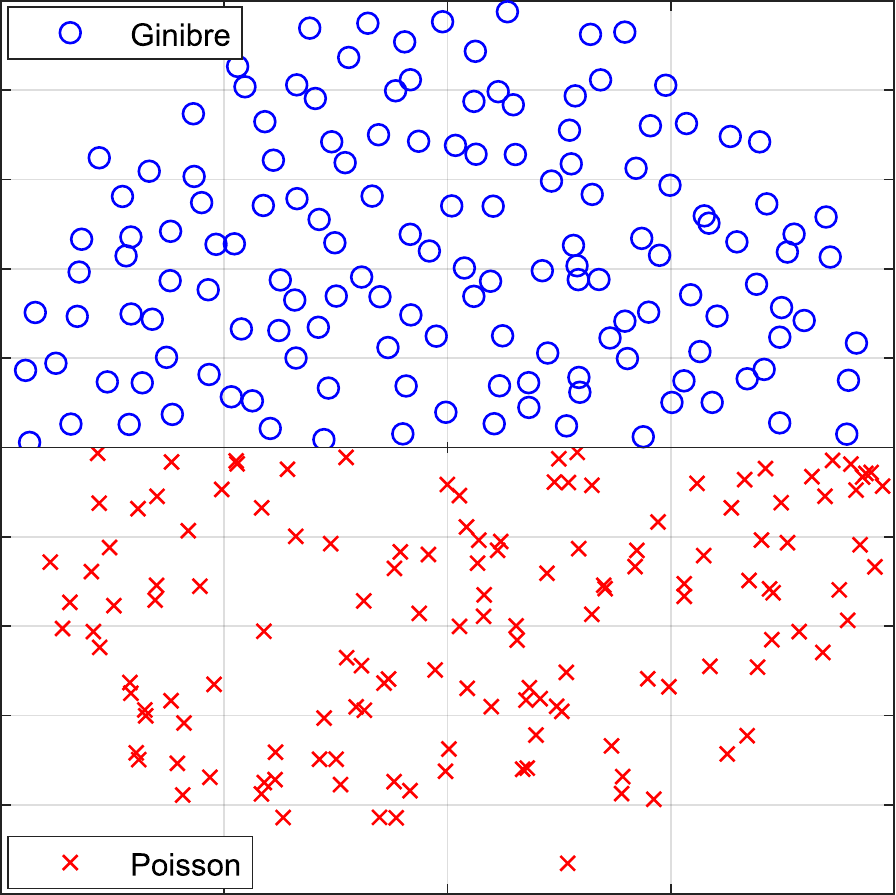}
  \caption{Visual comparison of nest locations of the Common buzzard
    with a typical snapshot of random matrix eigenvalues.  \newline
    \textbf{Left:} Map showing the locations of $251$ observed 
    buzzard nests from the year $2019$ north-west of Bielefeld. Note how
    smaller cities like Halle and Werther coincide with holes in the
    data. There are more nests outside the cluster shown here, but the
    positions of these have not been recorded. \newline
    \textbf{Right:} Points distributed according to the 2D Poisson
    distribution (bottom, red crosses) and eigenvalues of a large
    complex Ginibre matrix (top, blue circles). Both ensembles consist
    of $300$ points, the Poisson variables are generated uniformly on
    the unit disk, while the Ginibre variables are the eigenvalues of
    a complex Gaussian random matrix normalised to the unit disk. Only
    the top/bottom half disk is shown for an easier comparison. Notice
    how the Poisson points tend to cluster, whereas points from the
    Ginibre ensemble are more evenly spaced. This is reflected in the
    nearest neighbour spacing distributions \eqref{Eq:Poisson-2d} and
    \eqref{GinibreSpacing}, which we compare quantitatively to the
    buzzard nests below.}
\label{Fig:PositionPlot}
\end{figure}
\end{center}

All forest patches were visited in late winter to look for breeding
pairs and all nests of four raptor species (Common buzzard, Goshawk
Accipiter gentilis, Red Kite Milvus milvus and Honey Buzzard Pernis
apivorus) were recorded in either large-scale maps or GPS devices to
examine spatial distribution and interspecific competition. When an
incubating Common buzzard was observed on the nest in both March and
April, that pair was classified as breeding (in contrast to
non-breeders which did not occupy a nest). An illustration is shown in
the left panel of Figure~\ref{Fig:PositionPlot}.  \smallskip

Figure~\ref{Fig:PositionPlot} shows the nest locations for a single
year $2019$, where the population density is among the highest over
the observed period.  The effect of towns and smaller cities appears
to be visible as holes in the data set. The statistics on the
potential edge effects is too small to draw conclusions. We thus chose
to treat all the data points as being part of a bulk, without
considering edge effects. Notice that, within the Ginibre ensemble of
random matrices, the statistics at the edge of the spectrum (unit
disk) differs from that of the bulk points. There, the transition from
edge to bulk is very rapid though, within a distance of order
$1/\sqrt{N}$ from the edge for $N$ points; compare \cite{TaoVu} for
the mathematical discussion of this class of random matrices.

\begin{center}
\begin{figure}[ht]
  \textbf{2019 alone} \hspace{0.18\linewidth}
  \textbf{2015-2019} \hspace{0.18\linewidth}
  \textbf{All Years} \hfill\\
  \begin{turn}{90}\hspace{22pt}\textbf{Nearest Neighbour}\end{turn}
   \includegraphics[width=0.30\linewidth,angle=0]{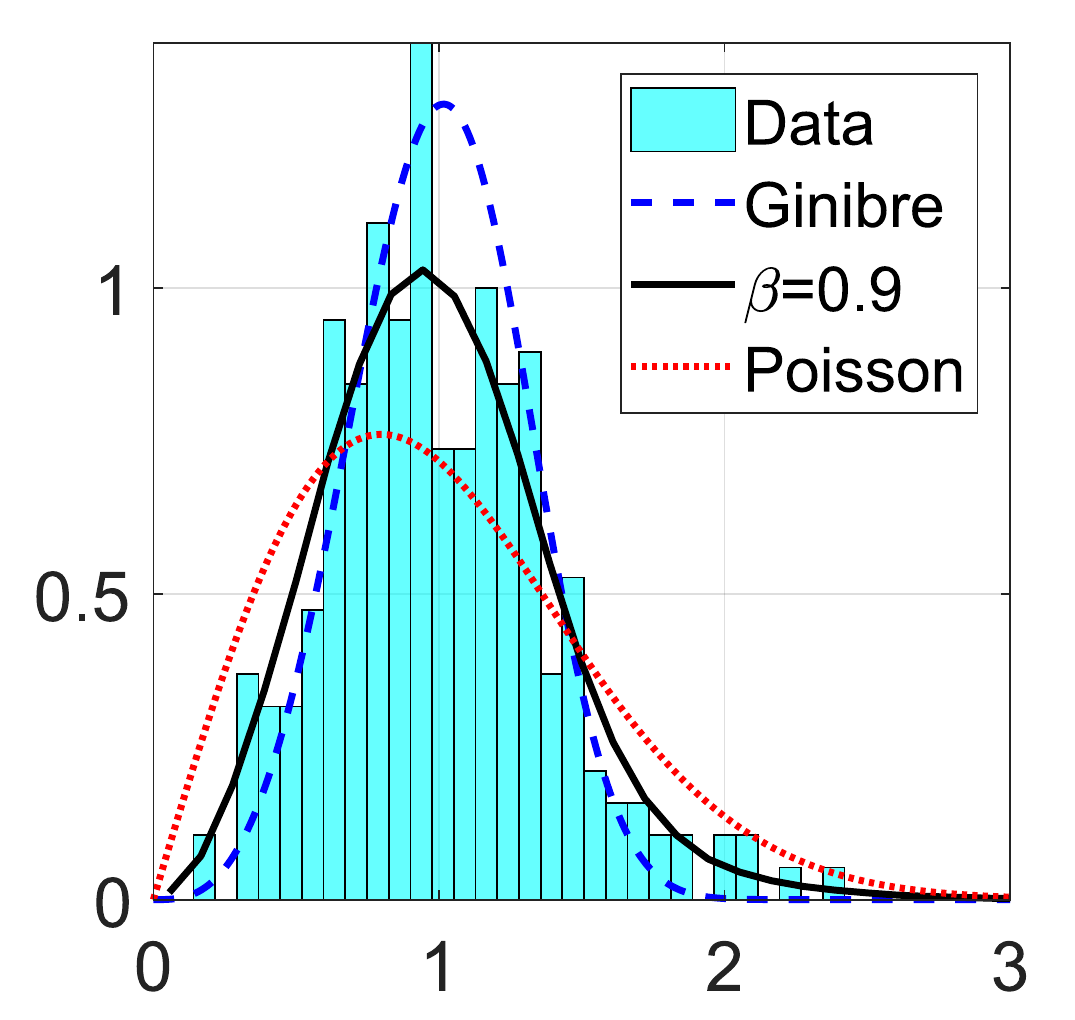}
   \includegraphics[width=0.30\linewidth,angle=0]
     {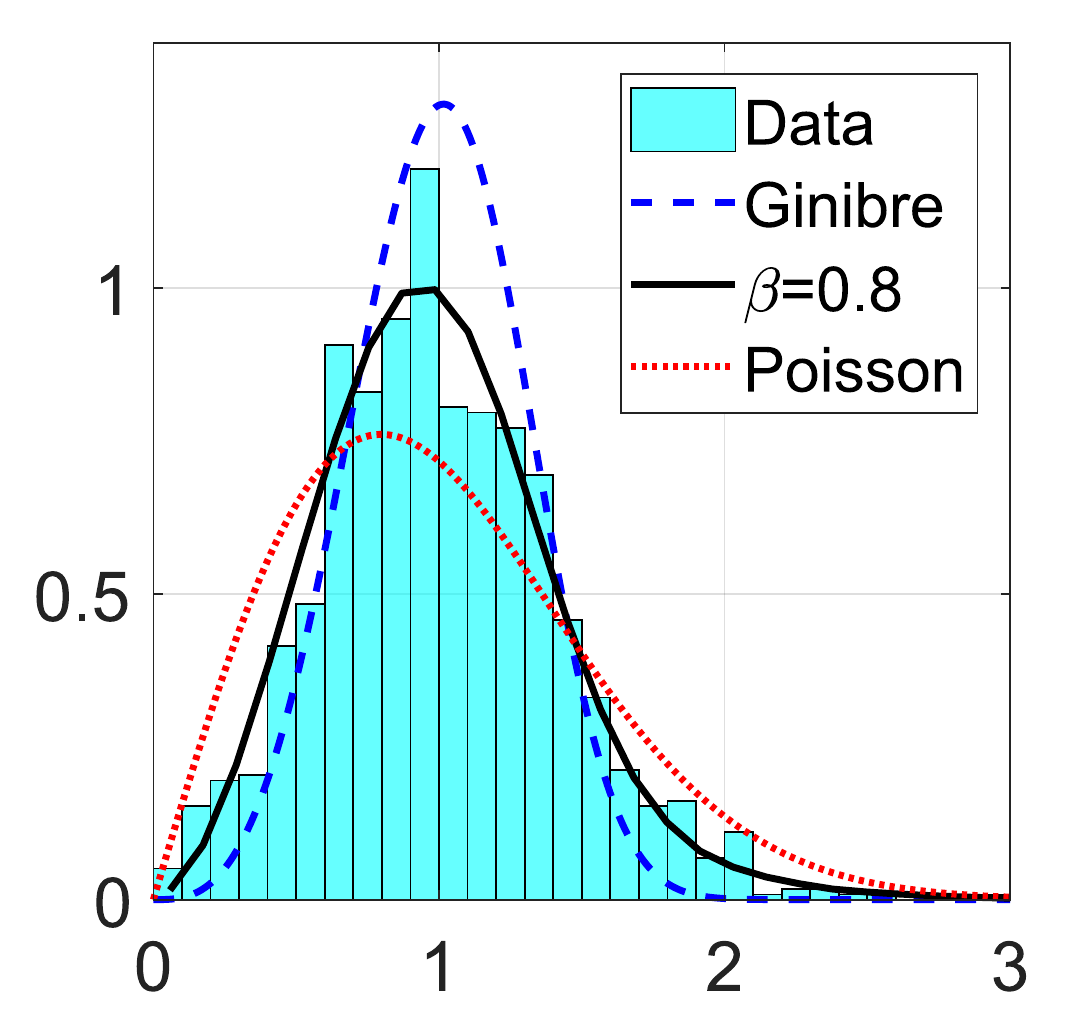}
   \includegraphics[width=0.30\linewidth,angle=0]
     {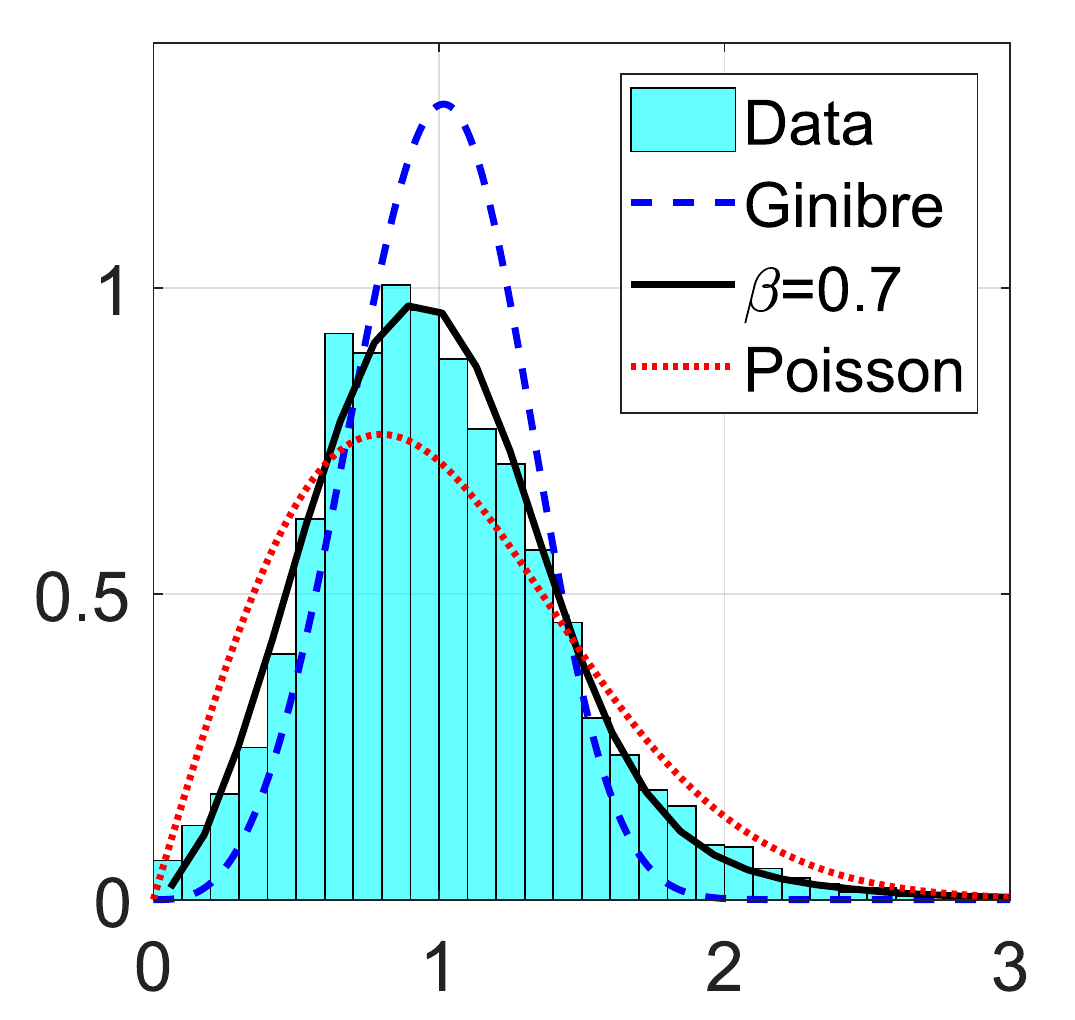}\\
\begin{turn}{90}\hspace{25pt}\textbf{Next-to-Nearest}\end{turn}
   \includegraphics[width=0.30\linewidth,angle=0]
     {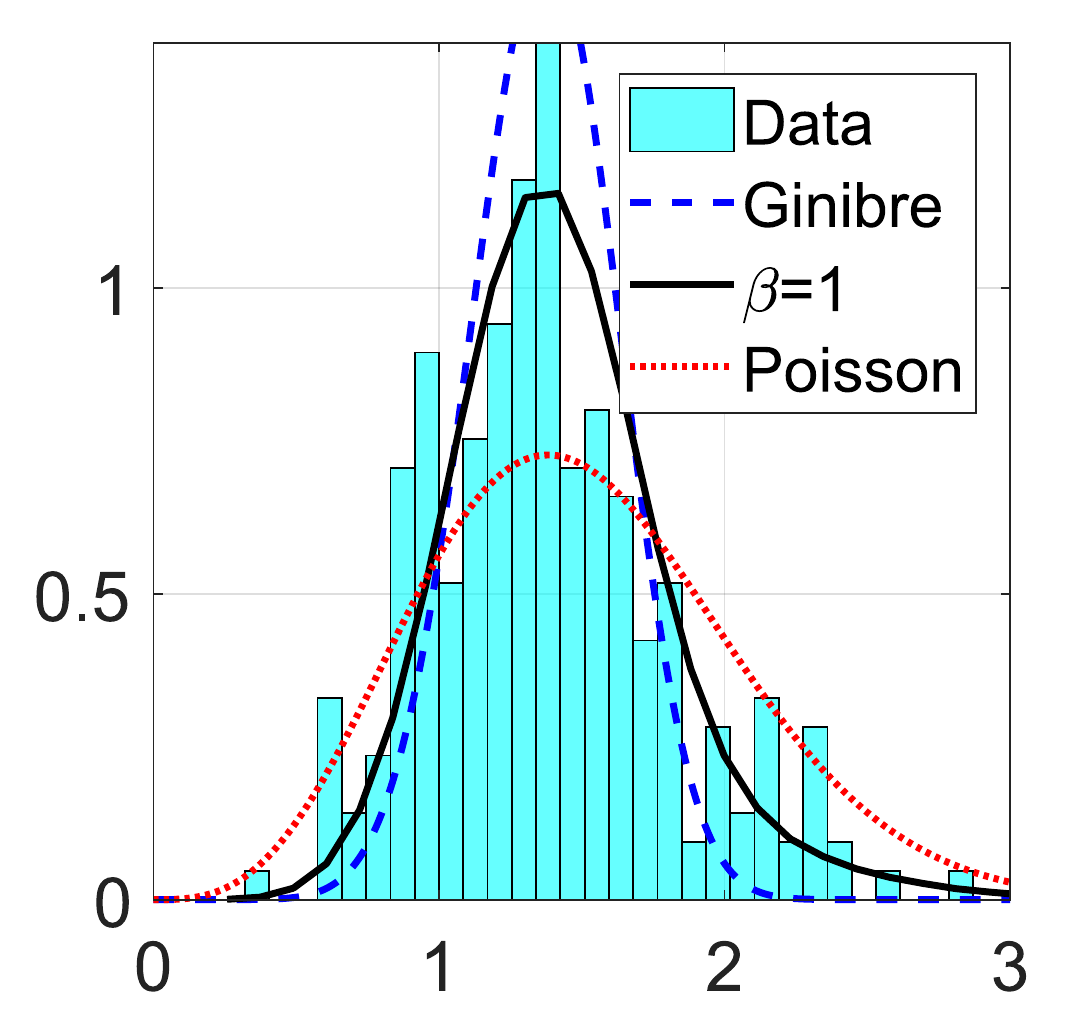}
   \includegraphics[width=0.30\linewidth,angle=0]
     {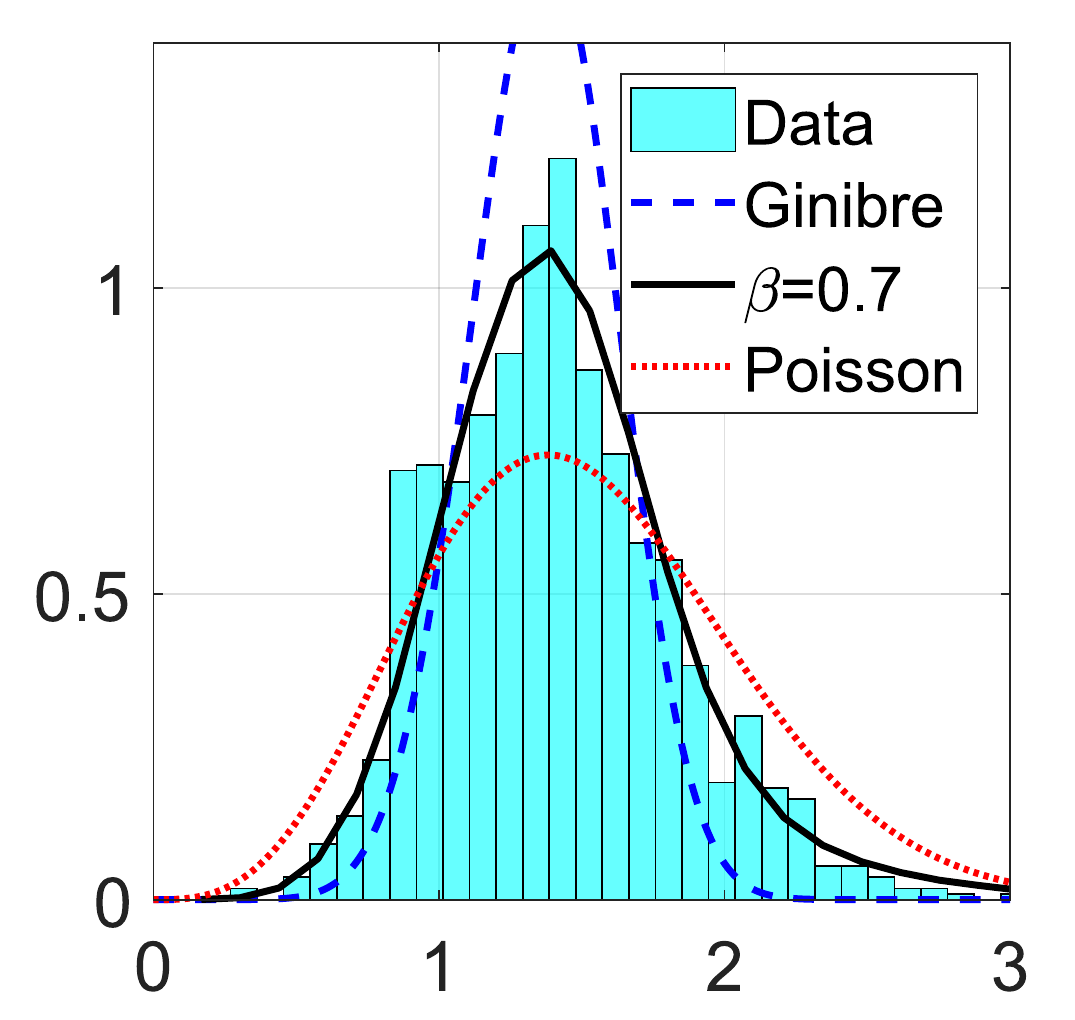}
   \includegraphics[width=0.30\linewidth,angle=0]
     {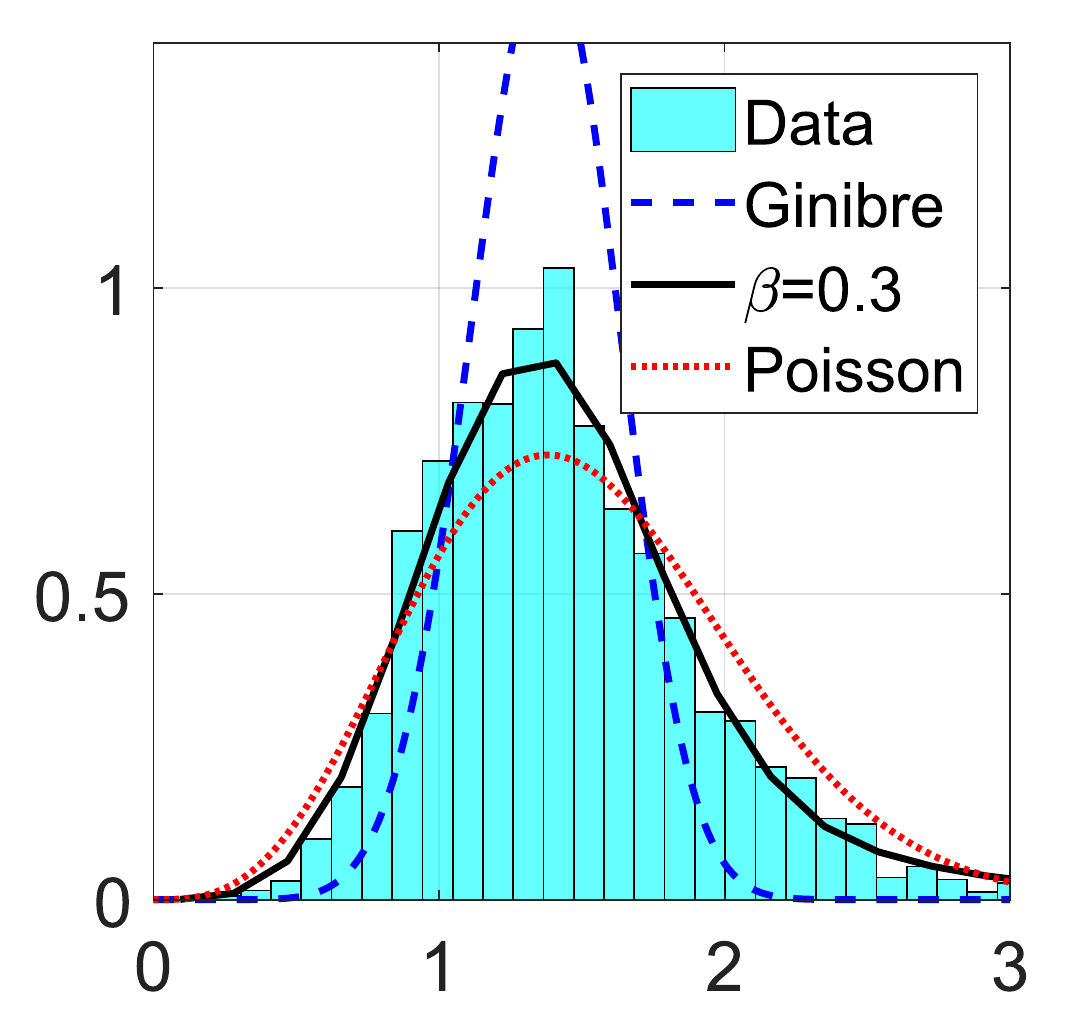}
\caption{Comparison of distances between nests to those of a 2D
       Coulomb gas \eqref{Eq:Coulomb} with a Kolmogorov fit, both
       nearest neighbour (NN, top row) and next-to-nearest neighbour
       (NNN, bottom row), where the repulsion $\beta$ is treated as a
       fitting parameter. We have also added the curves corresponding
       to the analytic expressions for Poisson and Ginibre variables;
       see Eqs.~\eqref{Eq:Poisson-2d} and \eqref{GinibreSpacing} for
       the top row, and Eqs.~\eqref{Eq:Poisson-2dNNN} and
       \eqref{GinibreSpacingNNN} for the bottom row. To illustrate our
       grouping of several years {explained in the main text}, we
       provide the fit of a single year (left column), of $5$ years
       (middle column), and of all $20$ years (right column).  Using
       one year, the fit quality is rather poor, but when using all
       years, we lose the temporal information. We therefore make the
       compromise of moving averages of $5$ years. The result is shown
       in Figure~\ref{Fig:PopulationBetaYears}. See also
       Figure~\ref{Fig:Grouping} in Appendix~\ref{C:Fit} for the
       effects of different group sizes.}
\label{Fig:CoulombFits}
\end{figure}
\end{center}

%%%%%%%%%%%%%%%%%%%%%%%%%%%%%%%%%%%

\section{Territorial Behaviour of Common Buzzards}\label{Sec:beta}

\begin{center}
\begin{figure}[ht]
  \includegraphics[width=0.49\linewidth,angle=0]
     {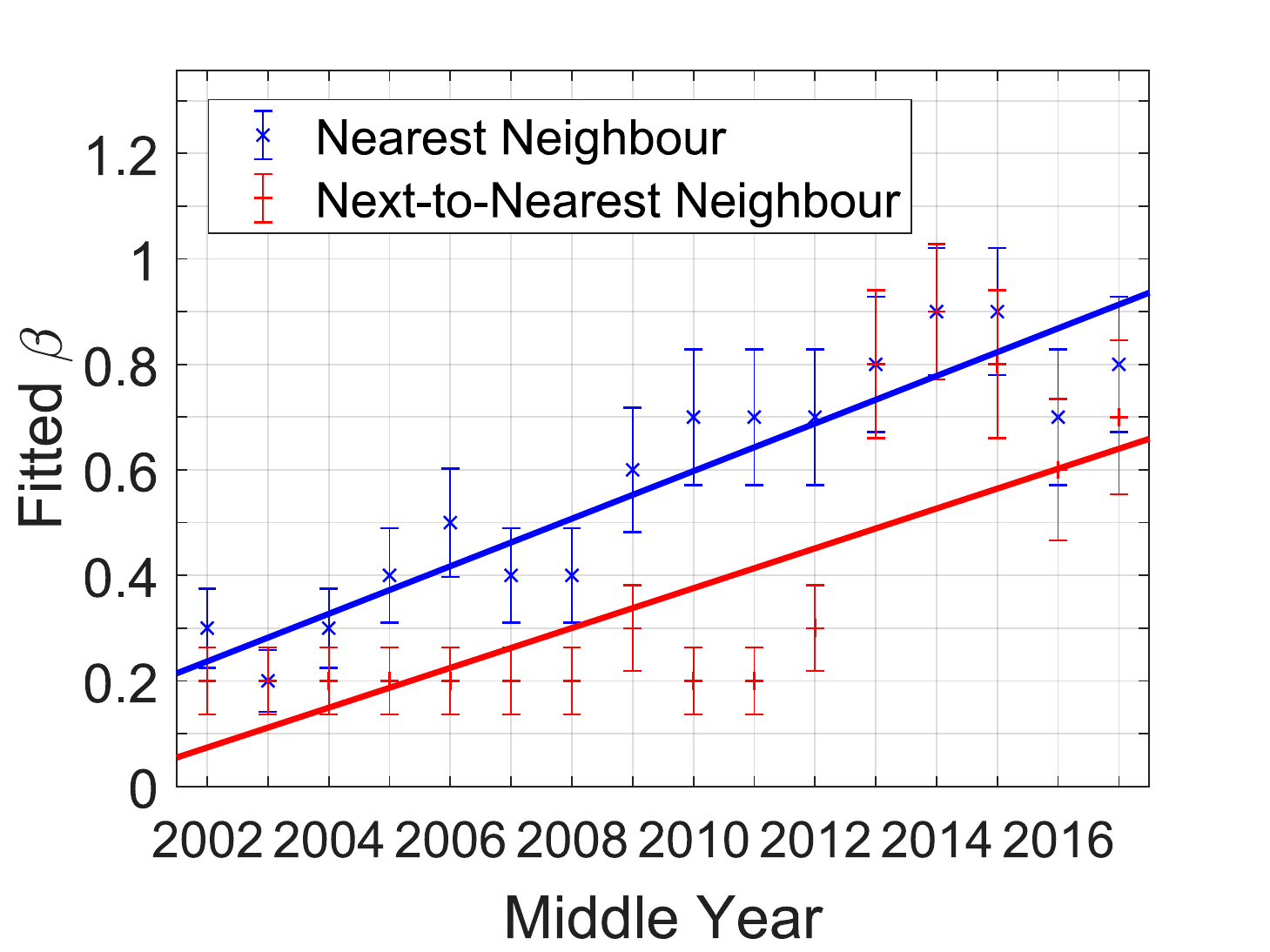}
  \includegraphics[width=0.49\linewidth,angle=0]
     {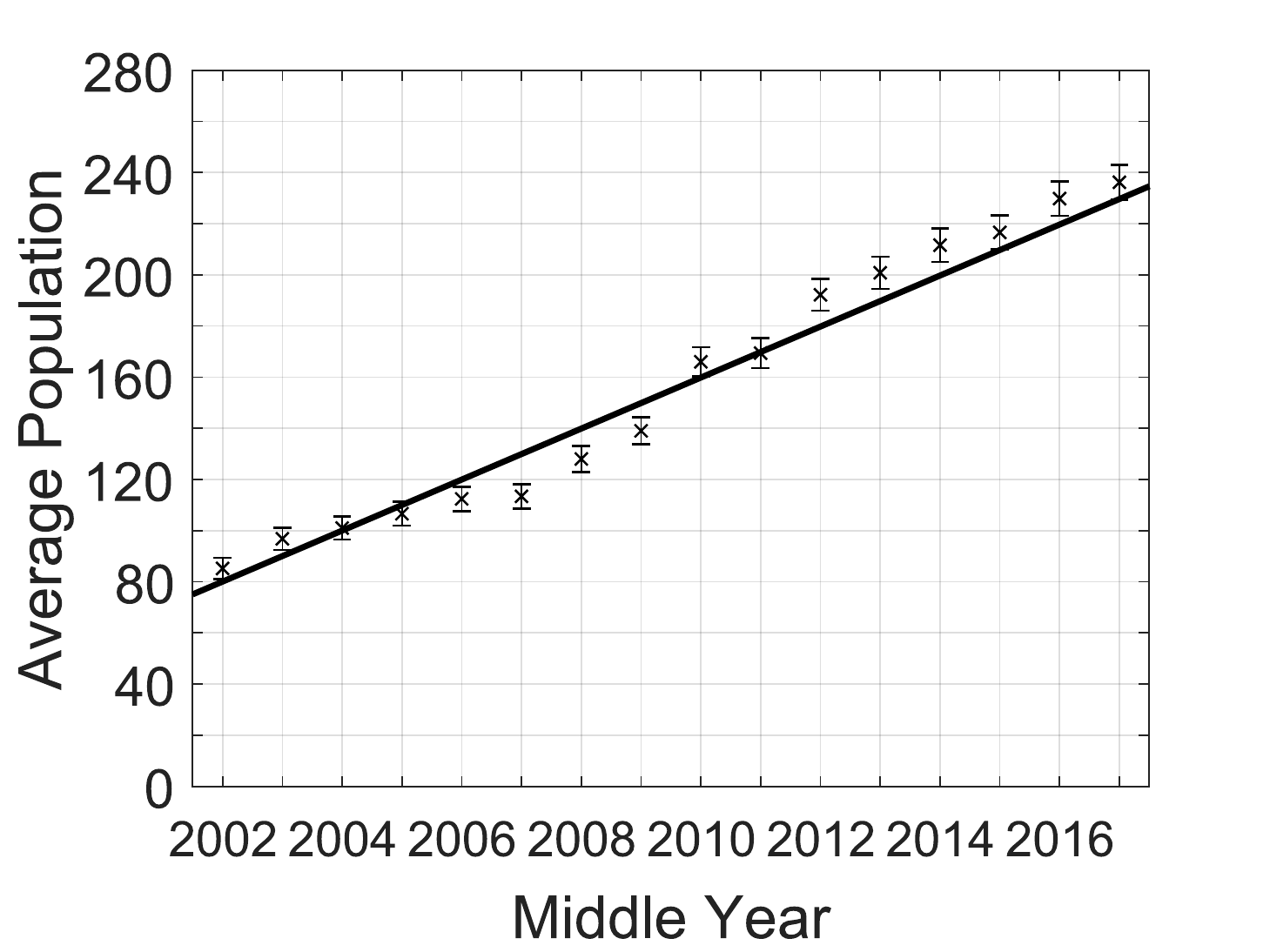}
\caption{\textbf{Left:} Comparison of the NN spacing distribution
       of all collected nests $2000-2019$ to a Kolmogorov fit
       for the effective parameter $\beta$ of the spacing
       distribution, obtained from a 2D Coulomb gas
       \eqref{Eq:Coulomb}; see Appendix~\ref{Sec:Coulomb}.  The fits
       to nearest (top blue data points) and NNN spacing (bottom red
       data points) are made in groups of $5$ years. That is, the
       first group is $2000 - 2004$ and the last is
       $2015 - 2019$; compare with Figure~\ref{Fig:Grouping} in
       Appendix~\ref{C:Fit} discussing different groups sizes.
       \newline As detailed in Appendix~\ref{C:Fit}, the statistical
       error bars are found by fitting the parameter $\beta$ in the
       spacing distribution obtained numerically from the Coulomb gas
       \eqref{Eq:Coulomb} to themselves (bootstrapping), that is, the
       spread of the $\beta$ obtained by fitting each of the $10^4$
       realisations to their average distribution. To avoid any
       influence from the choice of bin width for our data, we have in
       fact fitted to the cumulative distribution of the spacing
       distributions, using the Kolmogorov distance. A least-squares
       fit with a straight line is provided for both spacing
       distributions, to show the trend.  \newline \textbf{Right:} The
       averages of the population size over the same groups of $5$ years
       as well as a fit with a straight line. The errors of the
       individual years are here assumed Poissonian and propagated to
       the averages.  Note that the linear increase in size coincides
       with the increase in $\beta$.}
\label{Fig:PopulationBetaYears}
\end{figure}
\end{center}

Ideally, we wish to investigate the territorial behaviour over the
years. Unfortunately, the individual years have too few data points
for comparison, as typically of the order of $1000$ data points would
be needed to make a good fit; see Figure~\ref{Fig:CoulombFits} (left)
and Figure~\ref{Fig:Grouping} (left) in Appendix \ref{C:Fit}.  We
therefore group the nests in ensembles of $5$ years
($2000 \minu 2004$, $2002 \minu 2005$, ...) as follows.  For each
year, the spacings are determined individually and then put into an
ensemble. After unfolding the data points as explained in
Appendix~\ref{B:unfold}, we fit the nearest and next-to-nearest
neighbour spacings {of every ensemble} to a Coulomb gas, with $\beta$
as the fitting parameter describing the local repulsion; see
Figures~\ref{Fig:CoulombFits} and
\ref{Fig:PopulationBetaYears}. Notably, these fits to NN and NNN do
not give the same value for $\beta$ within the same ensemble.  Let us
first compare the $\beta$-dependence within each ensemble. Initially,
the next-to-nearest neighbour spacing gives a significantly lower
value for $\beta$, which is about half of that of the nearest
neighbour $\beta$-value.  At first, it is close to the Poisson process
at $\beta=0$. This suggests that the correlation length is relatively
small: the buzzards are aware of their direct neighbours, but do not
have long-range interactions. This differs from the long-range
interaction of the Coulomb force.

Next, we also compare the growth in population in
Figure~\ref{Fig:PopulationBetaYears} (right) to the time dependence of
the repulsion measured in $\beta$ in
Figure~\ref{Fig:PopulationBetaYears} (left) over the years and find
that an increase in population coincides with an increase in
repulsion.  This is indicated by the linear fits in
Figure~\ref{Fig:PopulationBetaYears}. Note that the observed area is
roughly the same through the years, so the population is proportional
to the density.  Surprisingly, beyond a certain critical density, the
$\beta$-value of both spacings becomes comparable, indicating an
increase in correlation length.  For Coulomb gases, the normalised
spacing distribution is invariant under a change in density, but here
the added population makes the Common buzzards more territorial. This
suggests that they care only about their closest neighbours, but not
about the overall scale of the environment. In other words, there
exists a length scale that is not present in the global potential of
the Coulomb gas.

%%%%%%%%%%%%%%%%%%%%%%%%%%%%%%%%%%

\section{Conclusion and Open Questions}\label{Sec:conc}

Above, we compared the spacing distributions of the nests of the Common
buzzard in an area of the Teutoburger Wald to a one-parameter family
of correlated random variables that includes the behaviour of Poisson
random variables and of complex eigenvalues of random matrices in
limiting cases. We find that it provides a good effective description
of the repulsion among neighbouring nests, thus quantifying the
territorial interaction between the birds.  This allows us to isolate
population density effects over time, where we find an increase in
repulsion through an increase of absolute population density, and to
also gauge the correlation length of the interactions. We observe that
the Common buzzard seems to care more about nearest neighbours, while
next-to-nearest neighbours only become important beyond a certain
population density.

One of the open questions is to come up with a simple phenomenological
model that would lead to similar statistics as the one-parameter
family of a 2D Coulomb gas that we found to be effective in our data
analysis.  In a more theoretical direction, the explicit data analysis
has shown that the theoretical results from \cite{AKMP} needed to be
extended in several directions, to include longer range correlations
and to efficiently deal with small data sets. This will certainly
inspire future activities on the mathematical side as well, for
instance to include different charges to model several species.
\bigskip

\textit{Acknowledgements:}
This paper grew out of a collaboration between members of three
distinct research groups, all supported by the German Science
Foundation (DFG), namely CRC1283 ``Taming uncertainty and profiting
from randomness and low regularity in analysis, stochastics and their
applications'' (GA and MB), IRTG2235 ``Searching for the regular in
the irregular: Analysis of singular and random systems'' (AM), and
CRC/Transregio 212 ``A Novel Synthesis of Individualisation across
Behaviour, Ecology and Evolution: Niche Choice, Niche Conformance,
Niche Construction (NC)${}^3$" (OK).  We thank Ellen Baake for useful
discussions and careful comments on the manuscript, as well as the
School of Mathematics and Statistics of the University of Melbourne
for hospitality (G.A.). \smallskip

\begin{appendix}
%%%%%%%%%%%%%%%%%%%%%%%%%%%%%%%%%%%%

\section{Spacing Distributions from Poisson, Ginibre and Coulomb 
     Gas}\label{Sec:Coulomb}
	
   Here, we recall the spacing distributions between nearest and
   next-to-nearest neighbours of correlated and uncorrelated random
   variables and discuss their relevance, related to the concept of
   universality in random matrix theory.

Let us begin with the uncorrelated case, the Poisson distribution
(Poi).  Given a set of uniformly distributed, uncorrelated points
in 2D, the distances between nearest neighbours (NNs) and
next-to-nearest neighbours (NNNs) follow the respective spacing
distributions
\begin{align}\label{Eq:Poisson-2d}
   p_{\mathrm{Poi}}^{\mathrm{(NN)}}(s) \; & = 
    \; \frac{\pi}{2} s\ \e^{-\pi s^2/4} ,\\[1mm]
   p_{\mathrm{Poi}}^{\mathrm{(NNN)}}(s) \; & =
   \; \frac{\pi^2}{8} s^3\ \e^{-\pi s^2/4},
   \label{Eq:Poisson-2dNNN}
\end{align}
in the limit of large data sets. Here, both the zeroth and first
moments are normalised to 1.  Note that in contrast to 1D, where the
Poisson distribution for NN is the simple exponential $\e^{-s}$,
without repulsion among NN, uncorrelated points in 2D also seem to
repel each other, with the linear factor stemming from the 2D area
measure. The derivation of Eq. \eqref{Eq:Poisson-2d} is standard.
Computing the probability that the NNs of a given point lie in a ring
of radius $s$ and thickness d$s$, with all other points being outside
that ring, leads to \eqref{Eq:Poisson-2d} in the limit of large
particle numbers.  The computation of NNN in \eqref{Eq:Poisson-2dNNN}
merely uses independence and \eqref{Eq:Poisson-2d}, where we refer to
Appendix A in \cite{Tomasetal} for a derivation in arbitrary
dimensions, including higher-order neighbours.

Let us turn to the spatial statistics based on random matrices.  They
have been conceived as a null model, for instance in the analysis of
multiple time series.  In that case, the data matrix with several
measured times series of equal length as rows is replaced by a random
matrix with independent Gaussian entries, with zero mean and unit
variance.  Taking this matrix times its transposed we obtain a
positive definite matrix, the Wishart matrix, and precise analytic
statements about the eigenvalues of this matrix can be made
\cite{For}. The eigenvalues become strongly coupled, due to the Jacobi
determinant from the transformation that diagonalises the Wishart
matrix.  In the limit of large matrix size, these results become
universal, that is, they hold for a much larger class of random matrix
entries.  In many instances, the majority of eigenvalues obtained from
the matrix of time series is well described by those of the random
matrix, after proper rescaling.  Further applications of the universal
random matrix predictions to data that are not necessarily eigenvalues
of a matrix or an operator are mentioned in the Introduction.

In order study complex eigenvalues, we consider $N\times N$ complex,
non-Hermitian random matrices, with independent entries drawn from the
complex normal distribution, the complex Ginibre (Gin) ensemble
\cite{Ginibre}. We directly consider the random matrix now, without
multiplying by its Hermitian conjugate, and thus its eigenvalues are
complex.  As for the real eigenvalues above, the complex eigenvalues
of such matrices become correlated random variables, due to the Jacobi
determinant of the Schur decomposition.  The joint density of complex
eigenvalues reads \cite{Ginibre}
\begin{equation}\label{Eq:Coulomb}
   p^{}_{\mathrm{Cou}, \beta} 
   (z^{}_1, \ldots , z^{}_N) \; \propto \; 
   \exp \biggl( \beta\sum_{1\leq j<k\leq N}
   \log |z^{}_k - z^{}_j |-\sum_{\ell=1}^{N} | z^{}_{\ell} |^2 \biggr) ,
\end{equation}
with $\beta=2$.  It equals the static gas of $N$ charged particles at
the locations of the complex eigenvalues, repelling each other with
respect to the long-range 2D Coulomb (Cou) interaction at a specific
inverse temperature $\beta=(k_BT)^{-1}$, with $k_B$ the Boltzmann
constant.  The particles (eigenvalues) are kept together through a
confining, quadratic potential $V(z)=|z|^2$, that originates from the
Gaussian distribution of matrix elements in the Ginibre ensemble.
Compared to standard conventions for the Gibbs measure in statistical
mechanics, we have rescaled the positions as $\beta|z_j|^2\to|z_j|^2$,
allowing us to take the limit $\beta\to0$ below.

The limiting NN and NNN spacing distributions among complex
eigenvalues with joint distribution \eqref{Eq:Coulomb} at $\beta=2$
can be derived explicitly in the large-$N$ limit:
\begin{align}\label{GinibreSpacing}
	p^{\mathrm{(NN)}}_{\mathrm{Gin}}(s) \; & = \;
	\left(\prod_{k=1}^{\infty}\frac{\Gamma(1 + k, s^2)}{k!}\right)
	\sum_{j=1}^{\infty} \frac{2s^{2j+1}\e^{-s^2}}{\Gamma(1 + j, s^2)},
	\\[2mm]
       p^{\mathrm{(NNN)}}_{\mathrm{Gin}}(s) \; & = \;
	\left(\prod_{k=1}^{\infty}\frac{\Gamma(1 + k, s^2)}{k!}\right)
        \sum_{j=1}^{\infty}\sum_{\substack{k=1 \\ k \ne j}}^{\infty}
        \frac{\gamma(1+j, s^2)}{\Gamma(1 + j, s^2)}
	\frac{2s^{2k+1}\e^{-s^2}}{\Gamma(1 + k, s^2)} .
\label{GinibreSpacingNNN}
\end{align}
Here, $\Gamma(1 + k, s^2)=\int_{s^2}^\infty t^k \e^{-t}\dd t$ and
$\gamma(1 + k, s^2)=\int^{s^2}_0 t^k \e^{-t}\dd t$ are the incomplete
Gamma functions.  The exact expressions at finite matrix size, where
sums and products are truncated at $N-1$, converge very rapidly. The
derivation for NN \eqref{GinibreSpacing} from \cite{GHS1988} can be
sketched as follows.  Placing one eigenvalue by hand at the origin,
for instance by setting $z^{}_1=0$ in \eqref{Eq:Coulomb}, the gap
probability that all other eigenvalues are at least at radial distance
$s$ is obtained by integrating the joint density \eqref{Eq:Coulomb}
over all angles and over all radii larger than $s$.  Algebraic
manipulations lead to the gap probability, being equal to the product
in front of the sum in \eqref{GinibreSpacing}. A differentiation with
respect to $s$ then gives the spacing distribution. The same idea
leads to the NNN expression \eqref{GinibreSpacingNNN} or higher order
spacings.  The generating function for the corresponding gap
probability was calculated in \cite{APS}, and a short computation
leads to \eqref{GinibreSpacingNNN}.

Let us discuss some properties of the spacing distributions in
\eqref{GinibreSpacing} and \eqref{GinibreSpacingNNN}.  It is not
difficult to see that, for small arguments, the repulsion is much
stronger in the Ginibre ensemble, being proportional to $s^3$ for NN
in \eqref{GinibreSpacing}, and to $s^5$ for NNN in
\eqref{GinibreSpacingNNN}, compared to \eqref{Eq:Poisson-2d} and
\eqref{Eq:Poisson-2dNNN} in the Poisson case, respectively; compare
Figure~\ref{Fig:CoulombFits}.  For simplicity, we have not yet
rescaled the expressions in \eqref{GinibreSpacing} and
\eqref{GinibreSpacingNNN} by the respective first moments, as it is
done in Figure~\ref{Fig:CoulombFits} for comparison to data.  Both
distributions for NN and NNN in the Ginibre ensemble are universal, in
the sense that they hold beyond independent Gaussian distributions for
the matrix elements; see \cite{AHM,TaoVu} for the precise statements
of the theorems for different generalisations. Moreover, for all three
Ginibre ensembles with real, complex, or quaternion matrix elements,
the spacing distributions in the bulk all agree \cite{AKMP,BS}, which
is why we display a single distribution for the Ginibre ensemble in
Figure~\ref{Fig:CoulombFits}.  This is in sharp contrast to Hermitian
random matrices with real, complex or quaternionic matrix elements,
which lead to three different spacing distributions among the
respective real eigenvalues \cite{GMW}.

The point process interpolating between Poisson and Ginibre ensemble
is now simply given by the joint distribution \eqref{Eq:Coulomb}, when
allowing $\beta$ to vary between $\beta=0$ for Poisson and $\beta=2$
for Ginibre.  In this sense, the 2D Coulomb gas \eqref{Eq:Coulomb}
provides a one-parameter family that interpolates between uncorrelated
and random matrix behaviour of random variables in 2D.  In the
large-$N$ limit, the global density of particles converges, for all
$\beta>0$, to a constant function on the unit disk (the so-called
circular law), with height $1/\pi$ in our conventions for the area
measure; see the top right panel of Figure~\ref{Fig:PositionPlot}.
For local correlations such as the spacing distributions, the
universality of the 2D Coulomb gas for arbitrary $0<\beta\neq 2$ is an
open problem; see \cite{Sylvia} for a detailed account.

Since no closed formula is known for the spacing for general
$0<\beta\neq 2$, we compute the spacing distributions for NN and NNN
numerically, with an importance sampling Metropolis--Hastings
algorithm; see for instance \cite{Has,CF}.  We have generated a
library of distributions in steps of $0.1$ for $\beta\in(0,2)$ for our
fits; see \cite{AKMP} for further details {for NN}.  Perturbatively,
it is clear from \eqref{Eq:Coulomb} that {the repulsion of NN for
  general $\beta$} increases polynomially, that is, we have
$p^{\mathrm{(NN)}}_{\mathrm{Cou}, \beta}(s) \propto s^{\beta+1}$ as
$s \to 0$; see the top row of Figure~\ref{Fig:CoulombFits} for an
illustration of the different behaviour at the origin.

%%%%%%%%%%%%%%%%%%%%%%%%%%%%%%%%%%%

\section{Unfolding of the Spectrum}\label{B:unfold}

Unfolding is a procedure to remove system-specific properties from
spectral data, in order to extract local correlations that have a
chance to be universal. For 1D, there is a unique, standard way to
unfold, described for example in \cite[Sec.~3.2.1]{GMW}. In 2D,
uniqueness is lost; see \cite{Tilo} for the criteria unfolding has to
satisfy. Here, we will follow the approach established and
successfully tested for NN in \cite{AKMP} in the case of 2D data.

The general problem in $D$ dimensions can be stated as follows.  The
average spacing between two points close to a reference point $z^{}_0$ 
is proportional to the inverse $D$-th root of the density at that point,
$\rho(z^{}_0)^{-1/D}$. Unfolding is a map that normalises the density
(and thus the spacing distribution). It allows to remove the effect of
the average (av) or mean spectral density $\rho_{\mathrm{av}}(z)$, which 
is typically system-specific, and to separate it from the fluctuations
(fl) around this density,
\begin{equation}\label{Eq:rhosplit}
	\rho(z) \, = \,
    \rho^{}_{\mathrm{av}}(z)+ \rho^{}_{\mathrm{fl}}(z).
\end{equation}
These fluctuations can often be described by simple, universal models,
such as the predictions from Appendix~\ref{Sec:Coulomb}.  In our 2D
case, unfolding consists of a map of complex coordinates
\[
   z \, = \, x + \text{i}y \; \longmapsto \;
    z^\prime \, = \, x^\prime + \text{i}y^\prime
\]
to new coordinates, in which the new density is normalised,
$\rho^{}_{\mathrm{av}}(x^\prime,y^\prime)=1$.  Following \cite{AKMP},
we first approximate the average density $\rho^{}_{\mathrm{av}}(x,y)$
by a sum of smooth Gaussian distributions, as given in
\eqref{Eq:unfold}.  Unfolding is then obtained by multiplying the
distance of the NNs (or NNNs) to each point $z_i$ by the factor
$\sqrt{\rho^{}_{\mathrm{av}}(x^{}_i, y^{}_i )}$ in 2D. The resulting
$N$ unfolded spacings are collected, their density and first moment
normalised to 1, and then compared to the correspondingly normalised
distributions from Appendix~\ref{Sec:Coulomb}; see
Figures~\ref{Fig:CoulombFits} and \ref{Fig:PopulationBetaYears}.
Notice that for points generated according to the Poisson, Ginibre, or
Coulomb gas point process, the mean density is already flat for the
values of $N$ considered, see Figure~\ref{Fig:PositionPlot} (right),
and thus no unfolding is necessary here.

The approximate mean density $\rho^{}_{\mathrm{av}}(x,y)$ is obtained
by a sum of Gaussian distributions centred around each of the $N$ data
points $z_j$,
\begin{equation}\label{Eq:unfold}
  \rho^{}_{\mathrm{av}}(x,y) \; \approx \; \frac{1}{2\pi\sigma^2 N}
  \sum_{j=1}^N\exp\biggl( -\frac{\, | z-z_j |^2}{2\sigma^2} \biggr) ,
\end{equation}
where $\sigma$ is the width, which is initially a free parameter to be
chosen appropriately. In order to arrive at a smooth density
$\rho^{}_{\mathrm{av}}(x,y)$, $\sigma$ should be larger than the mean
spacing $\bar{s}$ between points.  In \cite{AKMP}, we tested this
approximation for NN for examples of random matrix ensembles where the
mean density is not flat and the local spacing distribution is known
to follow \eqref{GinibreSpacing}. There, the choice
$\sigma=4.5\bar{s}$ gave very good results, which is why we use the
same value for the approximation \eqref{Eq:unfold}, after determining
the mean spacing $\bar{s}$ for our data points for each individual
year.  The unfolded spacings for each year are then obtained by
multiplication with the corresponding mean density
$\sqrt{\rho^{}_{\mathrm{av}}(x^{}_i, y^{}_i )}$ at each data point
$z_j$.  These unfolded spacings are then put together in moving
windows of {ensembles of $5$ consecutive years}, which are chosen
large enough to have sufficiently many spacings (of the order of
$1000$) for a meaningful fit of the parameter $\beta$ of the Coulomb
gas \eqref{Eq:Coulomb}.

Let us emphasise that the unfolding \textit{removes} the trivial
effect through the increase of the global population density observed
in the period 2000--2019, and that the correlations among the unfolded
points represent local properties that characterise the presence (or
absence) of repulsion among data points.

%%%%%%%%%%%%%%%%%%%%%%%%%%%%%%%%%%%%%%

\section{Fitting Approach}\label{C:Fit}

We compare two distributions $f$ and $g$ with the Kolmogorov distance
\begin{equation}
   D^{}_{\mathrm{KS}} (F,G) \, := \, \| F - G \|^{}_{\infty}
\, = \, \sup_{x} \, \lvert F(x) - G(x) \rvert
\, \le \, 1 \, ,
\end{equation}
where $F$ and $G$ are the cumulative distributions of $f$ and $g$,
respectively.  This has the advantage of being unbinned and takes into
account that the distributions are normalised.

To estimate the error and to then make the linear fit of the
population size and the repulsion strength $\beta$, we do the
following.  For the population size, we assume the numbers are Poisson
distributed, where the width is the square root of the mean. The
uncertainty on the average populations in the right plot of
Figure~\ref{Fig:PopulationBetaYears} is found through error
propagation, though these are merely there to guide the eye. As the
grouped average population sizes are correlated, the linear fit is
made with the individual years and plotted on top; see \cite{Bar} for
more on error estimation and statistics. For the Coulomb gas fit, this
is non-trivial as the Kolmogorov fit does not give a clear connection
to the uncertainty the same way a least-squares fit does. Because of
the overlapping years in the moving average, the points are also
correlated, and error propagation is not clear here.

\begin{center}
\begin{figure}[ht]
  \includegraphics[width=0.49\linewidth,angle=0]
     {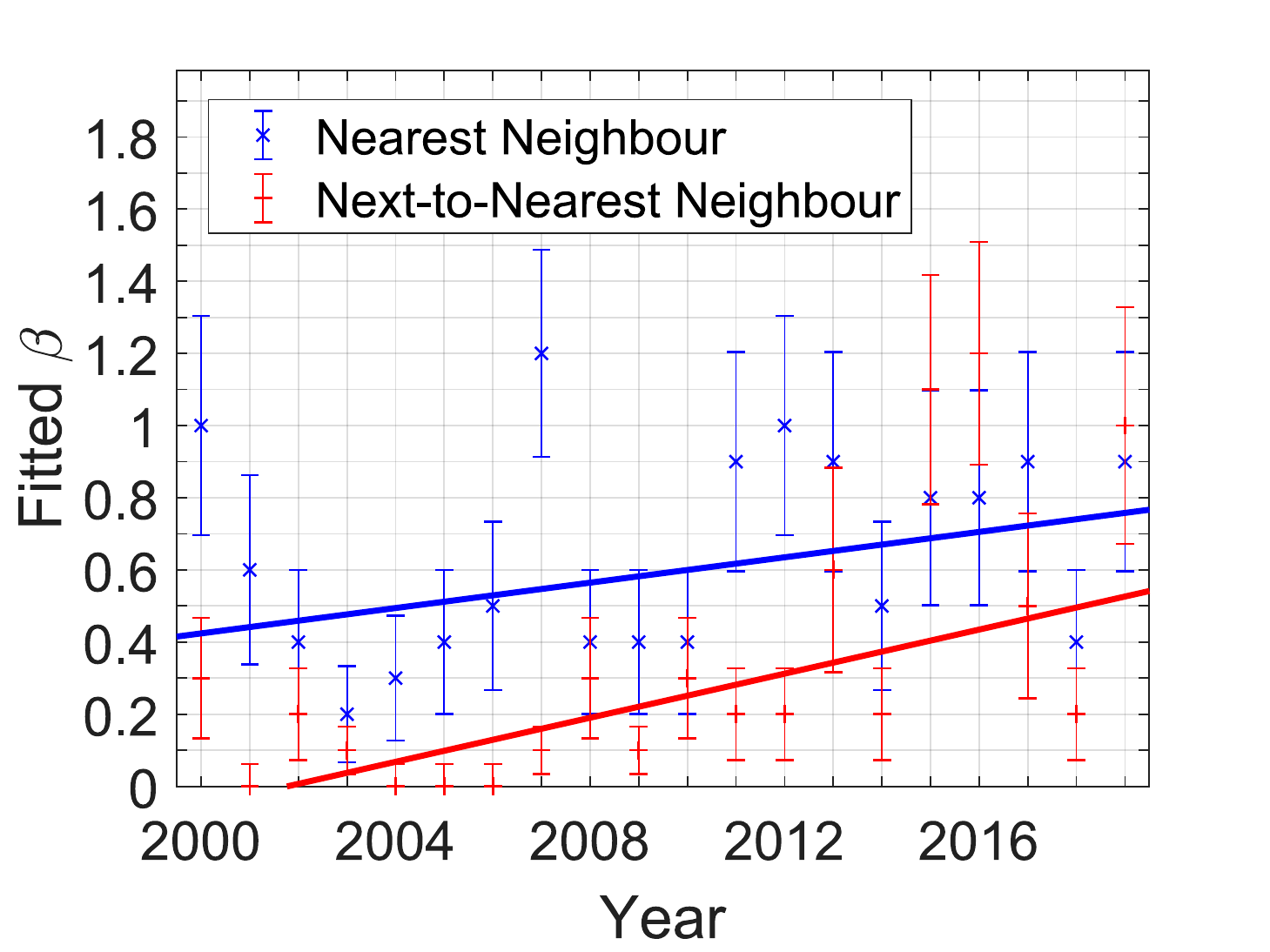}
  \includegraphics[width=0.49\linewidth,angle=0]
     {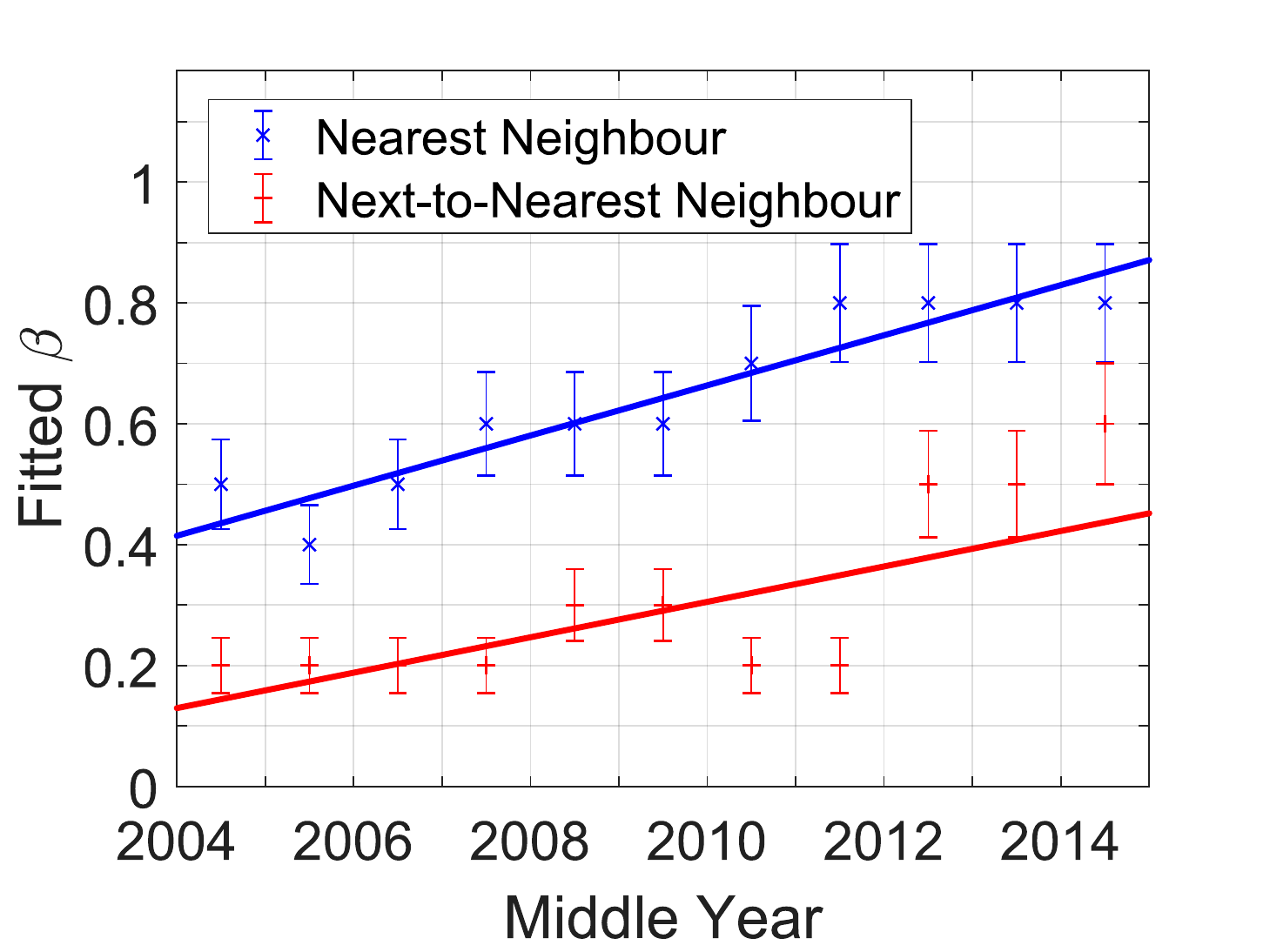}

\caption{The effects of different group sizes. Plotted is a
       comparison of the NN and the NNN spacing of all collected nests
       $2000 \minu 2019$ with a Kolmogorov fit for the effective parameter
       $\beta$ of the spacing distribution, obtained from a 2D Coulomb
       gas \eqref{Eq:Coulomb} in the same way as in
       Figure~\ref{Fig:PopulationBetaYears}, but with different
       numbers of years per group.\newline \textbf{Left:} No grouping,
       that is, each year is fitted on its own. The fluctuations that
       arise without grouping the years become apparent. \newline
       \textbf{Right:} Groups of 10 years. While the fluctuations are
       dampened, so is the change in $\beta$ for next-to-nearest
       neighbour spacing for later years. Parts of the temporal
       structure is lost in this way. For this reason, we choose 5
       years per group as a compromise.}
\label{Fig:Grouping}
\end{figure}
\end{center}

We therefore employ a method called \emph{bootstrapping}. We generate
a number of Coulomb gas realisations with a true $\beta=\beta_0$ and
make overlapping groups of 5 realisations the same way as we do with
the moving average of the nests. By fitting the $\beta$ of these
realisations, we can estimate the error of the fitting method. The
errors given in Figure~\ref{Fig:PopulationBetaYears} are the standard
deviations of the fitted $\beta$ for the corresponding
$\beta_0$. Because we group the realisations, we also have some
information about the cross-correlation between the years. We extract
this information by an average over groups of distance $k$ between the
midpoint for a given $\beta_0$ and use $V^{\beta_0}_k$ to denote the
correlation found here. We do not represent the nests completely with
this method, because we only group realisations of the same $\beta_0$,
but going into the individual $\beta_0$ of each year would defeat the
purpose of grouping. Instead, to compare two groups with different
$\beta_0$ and $\beta_1$, we use the heuristic combination
\begin{equation}
  V^{\beta_0,\beta_1}_k \; = \; 
   \sqrt{\frac{\left(V^{\beta_0}_k\right)^2
                            + \left(V^{\beta_1}_k\right)^2}{2}}\ .
\end{equation}
Here, the sum of squares reflects the structure of the
least-squares error, and the 2 in the denominator ensures that
$V^{\beta_0,\beta_0}_k = V^{\beta_0}_k$. From here, we may construct
the full covariance matrix of the grouped years in the left panel of
Figure~\ref{Fig:PopulationBetaYears}. The off-diagonal elements turn
out to be small compared to the diagonal elements, so the diagonal
part illustrated as error bars in Figure~\ref{Fig:PopulationBetaYears}
gives a reasonable idea about the uncertainty, but should not be taken
as a quantitative statement. The off-diagonal
elements are, however, still included in the linear fit.

How many years are included in each group of course influences the
results. We have chosen groups of $5$, because this is a compromise
where fluctuations are relatively small, but the temporal structure
still is visible. See Figure~\ref{Fig:Grouping} for the effect of
different choices of group size.
\end{appendix}

\end{document}